\documentclass[twocolumn,10pt,aps, pra, nofootinbib, superscriptaddress, showkeys,showpacs,longbibliography]{revtex4-1}
\usepackage{tikz}
\usepackage{graphicx}

\usepackage{epsfig}
\usepackage{dcolumn}
\usepackage{amsmath}
\usepackage{bm}
\usepackage{bbm}
\usepackage{ulem}
\usepackage{mathtools}
\usepackage{xcolor}
\usepackage{soul}
\usepackage{centernot}
\usepackage{dsfont}	
\usepackage{relsize}

\usepackage{epstopdf}

\usepackage{mathrsfs}
\usepackage{float}
\usepackage{caption}
\usepackage{subcaption}
\usepackage{ragged2e}
\usepackage{booktabs} % Top and bottom rules for table
\usepackage{amsfonts, amsmath, amsthm} % For math fonts, symbols and environments
\usepackage{wrapfig} % Allows wrapping text around tables and figures
\usepackage[font=small,labelfont=bf,justification=justified,format=plain]{caption}
\usepackage{color}
\usepackage{dsfont}
\usepackage{setspace}
\usepackage[colorlinks, citecolor = blue]{hyperref}
\definecolor{blue(pigment)}{rgb}{0.2, 0.2, 0.6}
\definecolor{dukeblue}{rgb}{0.0, 0.0, 0.61}
\definecolor{armygreen}{rgb}{0.29, 0.33, 0.13}
\definecolor{bulgarianrose}{rgb}{0.28, 0.02, 0.03}
\definecolor{coolblack}{rgb}{0.0, 0.18, 0.39}
\definecolor{burgundy}{rgb}{0.5, 0.0, 0.13}
\definecolor{darklava}{rgb}{0.8, 0.1, 0.4}

\makeatletter
\newcommand{\ovast}{\bBigg@{3}}
\newcommand{\vast}{\bBigg@{4}}
\newcommand{\Vast}{\bBigg@{5}}
\makeatother
\usepackage{mathtools}
\DeclarePairedDelimiter\bra{\langle}{\rvert}
\DeclarePairedDelimiter\ket{\lvert}{\rangle}
\DeclarePairedDelimiterX\braket[2]{\langle}{\rangle}{#1 \delimsize\vert #2}

\begin{document}

\title{Gravitationally induced entanglement dynamics between two quantum walkers}% Force line breaks with \\
%\thanks{A footnote}%

\author{Himanshu Badhani}%
 \email{himanshub@imsc.res.in}
 \affiliation{The Institute of Mathematical Sciences, CIT Campus, Taramani, Chennai- 600113, India}
\affiliation{Homi Bhabha National Institute, Training School Complex, Anushakti Nagar, Mumbai 400094, India}
\author{C. M. Chandrashekar}
\email{chandru@imsc.res.in}
\affiliation{The Institute of Mathematical Sciences, CIT Campus, Taramani, Chennai- 600113, India}
\affiliation{Homi Bhabha National Institute, Training School Complex, Anushakti Nagar, Mumbai 400094, India}

%====================================================================
\begin{abstract}
Quantum walk is a synonym for multi-path interference and faster spread of a particle in a superposition of position space. We study the effects of a quantum mechanical interaction modeled to mimic quantum mechanical gravitational interaction between the two states of the walkers. The study has been carried out to investigate the entanglement generation between the two quantum walkers that do not otherwise interact. We see that the states do in fact get entangled more and more as the quantum walks unfold, and there is an interesting dependence of entanglement generation on the mass of the two particles performing the walks. We also show the sensitivity of entanglement between the two walkers on the noise introduced in one of the walks.  The signature of quantum effects due to gravitational interactions highlights the potential role of quantum systems in probing the nature of gravity.
\end{abstract}
%===================================================================
\maketitle
\noindent
\section{Introduction} 
	One of the most elusive quests in theoretical physics for almost a century now has been the understanding of the quantum nature of gravity. Some of the most promising theories of our age that are attempting to answer this question, work at length scales that are far beyond our experimental capabilities. For this reason, there has been an interest to look for the signature of quantum gravity, both at the cosmological scales\,\cite{ashtekar1996large}, as well as in tabletop experimental setups. The later approaches attempt to probe the Planck scale length \cite{brun1,brun2}, as well as the quantum nature of gravity itself by exploiting the phenomena of quantum interference and quantum entanglement. With the experimental advancement over last decade reporting control over quantum systems, a revived interest proposes experiments using quantum interference to answer if the gravitational attraction between two masses is quantum mechanical\,\cite{bose2017, marletto2017}. The basic set-up for both the papers consists of two massive particles, each prepared in a superposition of two position states. The particles then evolve under their mutual gravitational interaction. The claim is that if we see entanglement generation between the particles, we must conclude that the interaction between the two particles, which in this case is only gravitational, must be quantum mechanical in nature. This is because a classical interaction cannot generate entanglement. Further, the model of gravitational interaction between the two objects in a superposition of two positions requires that the gravitational attraction itself be in a superposition of two different values. This argument, in fact, finds its roots in the interaction between Feynman and colleagues including Bondi, Bergmann, Wheeler, and others at the 1957 Chapel Hill Conference\,\cite{feynman2011}. While discussing whether gravity at all should be quantized, Feynman proposes an experiment in which a little ball of diameter 1{\it cm} is prepared in a quantum superposition of two states. This ball is then used to move another object gravitationally which should (according to Feynman) carry the information of the quantum amplitudes (which can be checked by performing interference experiments on the second ball). The argument is that if one can prove that the information of the quantum amplitudes can travel across a gravitational channel, the channel must be quantum mechanical; unless, of course, quantum mechanics fails at mass ranges where gravity starts getting significant. The recent proposals in \cite{bose2017} and \cite{marletto2017} improve on this thought experiment to find a witness (in form of entanglement entropy) for the quantum mechanical nature of gravity using an interference set-up.
	\\ \\
	In this paper, we generalize the given model of gravitational interaction to a system of two particles performing quantum walks. The discrete-time quantum walks, version studied in this work provide a more controlled way of handling interferences, engineer any arbitrary configuration of state in a superposition of position space\,\cite{giordani2019} and model the effect of noise in the dynamics. They have also been used to simulate various quantum systems including Dirac fields (see eg.\,\cite{meyer1996quantum,  ryan2005, chand2008phasetran, mallick2017, chandrashekar2011disordered}). So, studying them under the action of a ``gravitational" model can give us interesting insights into the effects of such interactions in quantum systems and help us to probe further investigations towards the quantum nature of gravity. In our work we explicitly see how entanglement between two particles of given masses changes with time. We also explore the relationship between the mass of the quantum particles and the entanglement generation due to the gravitational interaction between them.
	\\
	In section \ref{two} we provide a short introduction to quantum walks and in section \ref{three} we present our model of two interacting quantum walks. Section \ref{four} includes the results on the entanglement growth with time and mass between the two walks. Conclusions and discussions of the results have been  presented in section \ref{five} .
	%============
	%\subsection{\label{sec:level2}Quantum Walks}
	%=======================
	\section{Quantum walks}\label{two} 
\begin{figure}[!h]
		\begin{subfigure}[b]{0.2\textwidth}
			\includegraphics[scale=0.3]{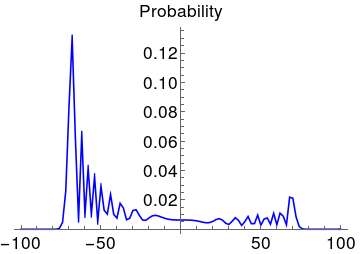}
			\caption{}
			\label{QW2Nonsymm}
		\end{subfigure} 	
		\hspace{2mm}
		\begin{subfigure}[b]{0.2\textwidth}	
			\includegraphics[scale=0.34]{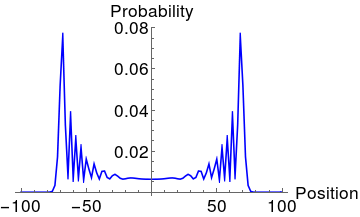}
			\caption{}
			\label{QW1symm}
		\end{subfigure}
		\caption{\small Probability distribution of the state after 100 steps of one dimensional quantum walk starting at the origin for the initial state (a) $\ket{\uparrow}$ and (b)  $(\ket{\uparrow}+i\ket{\downarrow})/\sqrt{2}$. The coin used is two dimensional Hadamard. Only the even positions are plotted, as the odd positions after even steps of the walk have zero probability.}
	\end{figure}

In a classical random walk, a particle hops over the different lattice points based on the result of a coin toss. In the same spirit, a quantum mechanical particle performs a quantum walk by hopping onto the different lattice sites based on the result of a quantum coin toss, a rotation in the space of internal degrees of freedom of the particle. The difference arises in the fact that unlike a classical coin that can exist exclusively in one or the other state, a quantum particle can exist in a superposition of two or more states. As a result, the hopping, which is conditional on the result of the coin toss, takes place in the superposition of two different lattice sites\,\cite{meyer1996quantum}. Thus the quantum property of superposition gives rise to the various constructive and destructive interference in the quantum walk which gives it a probability distribution that has properties drastically different from its classical counterpart. Here we present a standard description of the single particle quantum walk on a one dimensional position space.
	\\
	Let us consider a quantum particle with two internal state $\ket{\uparrow}$ and $\ket{\downarrow}$ at a localized initial position, the origin $\ket{x=0}\equiv\ket{0}$. When the particle is initially in only of the internal state, $\ket{\uparrow}$, we represent this combined state by $\ket{0}\otimes\ket{\uparrow}$, indicating that the state lives in the direct product Hilbert space of the spin (the coin space) and position Hilbert space, $\mathcal{H}_c$ and $\mathcal{H}_p$, respectively. Here we take our position states to be orthonormal. Orthogonality of the position states stems from the assumption that at each site the wave function has a spread of $\delta x$ which is much smaller than the lattice spacing $a$. So, the overlap of two different position states vanishes: $\braket{x}{y}=\delta_{x,y}$.\\
	The first step of the walk is to perform a rotation in the spin space (or a coin space), just like a coin toss, using a unitary operator, let us say a Hadamard operator $H_2 = \frac{1}{\sqrt{2}} \begin{bmatrix}
		1 & ~~1 \\
		1 & -1
	\end{bmatrix}$. This operation is represented as 
	\begin{equation}
		(I\otimes H_2 )\cdot (\ket{0}\otimes\ket{\uparrow})=\dfrac{\ket{0}\ket{\uparrow}+\ket{0}\ket{\downarrow}}{\sqrt{2}},
	\end{equation}
	where $I$ represents the identity operation on the position space. The coin operation is followed by the \textit{shift} operation, to effect a change of the state in position space, conditioned on the spin state. We can assign the shift to the right by one position for $\ket{\downarrow}$ states and to the left for $\ket{\uparrow}$. This operation, for any state at position $x$ will be of the form given by
	\begin{equation}\label{shift}
		S =\sum_{x}\ket{x-1}\bra{x}\otimes\ket{\uparrow}\bra{\uparrow}+\ket{x+1}\bra{x}\otimes\ket{\downarrow}\bra{\downarrow}.
	\end{equation}
	One step of the walk for a state $\ket{\psi}=\ket{0}\otimes\ket{\uparrow}$ will then be given by the operation $W\ket{\psi}=S\cdot(I\otimes H_2)\ket{\psi}$. $t$ steps of the walk will be realized by the $t$-times implementation of the walk operator, $W^t = [S \cdot(I\otimes H_2)]^t$.
	\\ \\
	Fig.\,\ref{QW2Nonsymm} shows the probability distribution after $t=100$ on the walker with initial state $\ket{0}\ket{\uparrow}$. We can see that the final distribution of a state depends on the kind of coin that is used and also on the initial state. For example, Fig.\,\ref{QW1symm} shows the distribution after 100 steps of the walk with Hadamard coin operation for the particle with the initial state $|0\rangle(\ket{\uparrow}+i\ket{\downarrow})/\sqrt{2}$. A more general form of coin operation,
	\begin{align}
		\label{ucoin}
		C(\theta) = \begin{bmatrix}
			~~\cos(\theta) &   \sin(\theta) \\
			-\sin(\theta) & \cos(\theta)
		\end{bmatrix}.
	\end{align}
	can be used in place of Hadamard operation $H_2$  to have more control over the walk dynamics. In this work we will use $(\theta)$ as the coin operation.
	\\
	Apart from the discrete time walks discussed here, quantum walks can also be defined for continuous time evolution\,\cite{farhi1998}. In addition, one can construct different forms of quantum walks by defining different combination of the coin and shift operators. One such example is the split-step quantum walk\,\cite{kitagawa2010} that has been shown to model the Dirac cellular automaton\,\cite{mallick2016dirac} and so is useful for quantum algorithms. 
	%==============================
	%\section{Quantum walks under mutual gravity}
	%===============================
	\\ \\
	\noindent
	\section{Quantum walks under mutual gravity}\label{three}
\begin{figure}[h]
		\centering
		\includegraphics[scale=0.45]{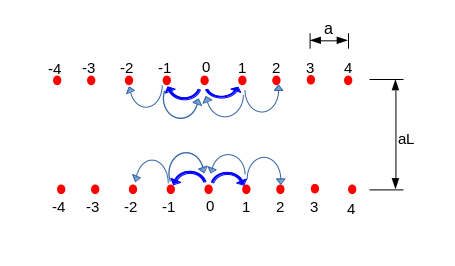}
		\caption{\small Two quantum walks on a plane, each restricted to a one dimensional walk parallel to each other. Distance between them is $L$ times the lattice distance $a$. We take $L$ to be large compared to the spread of the walk after $t$ amount of time.}
		\label{qwmodel}
	\end{figure}

\begin{figure}
		\begin{subfigure}[b]{0.2\textwidth}
			\includegraphics[scale=0.12]{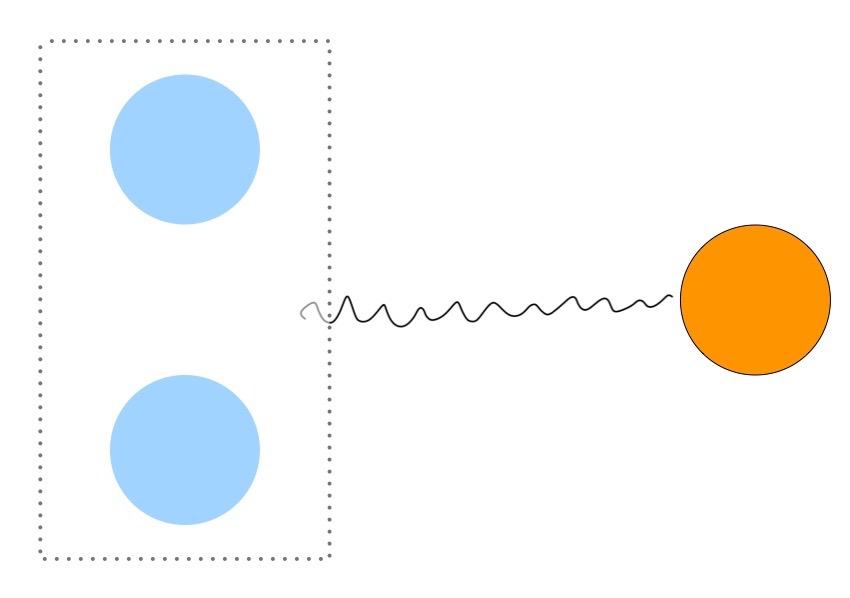}
			\caption{}
			\label{cgr}
		\end{subfigure} 
		\hspace{5mm}	
		\begin{subfigure}[b]{0.2\textwidth}	
			\includegraphics[scale=0.12]{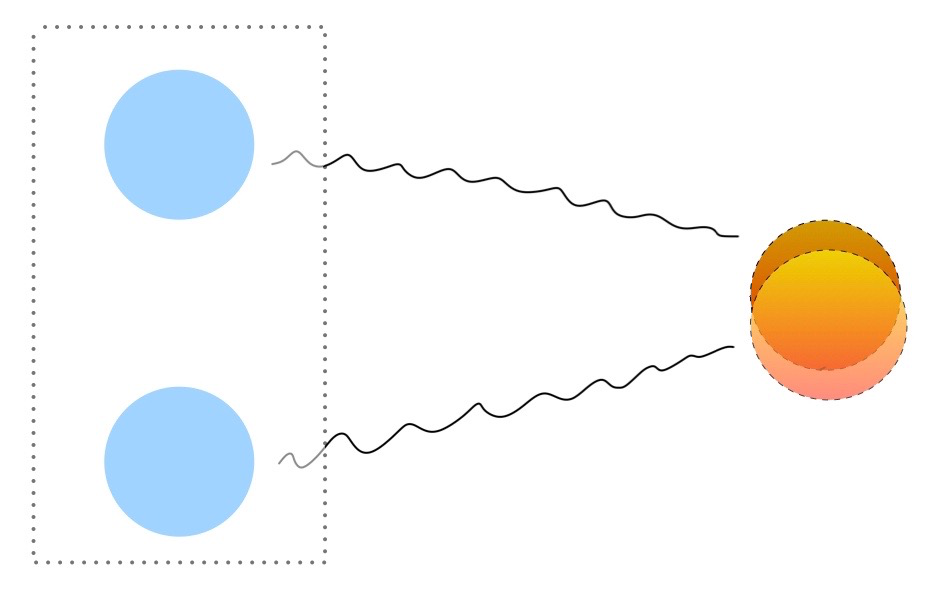}
			\caption{}
			\label{qgr}
		\end{subfigure}
		\caption{\small \textbf{Illustrating the difference between a classical and quantum field:} Given a particle in a superposition of two positions, shown by the blue disks inside a box, we consider two possible ways an interaction between this particle and another system can take place. Fig. \ref{cgr} shows the case when the particle produces a field in a singular configuration. Contrasting it with  Fig. \ref{qgr}, that shows the case when each component of the superposition produces its own field. The effective field in this can only be described as a superposition of two fields thus generated.}
		\label{modelsofgr}
	\end{figure}

	Here we will generalize the model of gravitational attraction proposed by Bose {\it et al.} \,\cite{bose2017}  and Martletto {\it et al.}\cite{marletto2017} to quantum walks. We consider a simplified system of two massive particles with two internal states in a two-dimensional discrete space, each performing discrete time quantum walks in one-dimensional space parallel to each other (see Fig.\,\ref{qwmodel}). We assume that the states are initially separable and the two walks start at the same time. After every step of the walk, the state evolves into a superposition of components at different positions. State of the first particle $\psi$ with two dimensional coin space, ($\ket{\uparrow}$ and $\ket{\downarrow}$) has the following general form after $t$ steps of walk, starting from the position $i=0$,
	\begin{equation}
		\psi(t)=\sum_{i=-t}^{t}\ket{i}\otimes(p^u_{i}(t)\ket{\uparrow}+p^d_{i}(t)\ket{\downarrow}).
	\end{equation}
	Here $i$ are the lattice indices on which the walk is being performed. $|p_i^u|^2$ ($|p_i^d|^2$) is the probability of spin state $\ket{\uparrow}$ ($\ket{\downarrow}$) at $i$.\\
	State of the second particle $\phi$ at position $j=0$ which starts a quantum walk parallel to first particle at time $t=0$ (see Fig.\,\ref{qwmodel}) has the following form at time $t$
	\begin{equation}
		\phi(t)=\sum_{j=-t}^{t}\ket{j}\otimes(q^u_{j}(t)\ket{\uparrow}+q^d_{j}(t)\ket{\downarrow}).
	\end{equation} 
	At time $t=0$ the composite state is $\ket{\psi\phi}$. If there are no quantum mechanical interactions between these states, the state of the system as a whole remains a product state. $\Psi (t) =\psi(t)\otimes\phi(t)$. 
	%\begin{equation}
	%\begin{aligned}
	%	\Psi (t)& =\sum_{i,j=-t}^{t} \ket{i}\otimes(p^u_{i}\ket{\uparrow}+p^d_{i}\ket{\downarrow})%\otimes \ket{j}\otimes(q^u_{j}\ket{\uparrow}+q^d_{j}\ket{\downarrow}).
	%\end{aligned}
	%\end{equation}
	\\ \\
	\noindent
	Let us introduce the interaction between the component $\ket{i_A}$ of the first walk and the component $\ket{j_B}$  of the second walk and ensure that it depends only on the distance between the two points, $i$ of the first lattice and $j$ of the second lattice, and not show any dependency on the other components or on the spin states. \color{black}With such an interaction, each component of the product state will evolve with respect to a different Hamiltonian. If this interaction is gravitational in the weak field limit, the component $\ket{i_A}\ket{j_B}$ will undergo unitary evolution with the Hamiltonian
	\begin{equation}\label{hamiltonian}
		\langle\hat{H}_{ij}\rangle=-\dfrac{Gm_Am_B}{|\hat{r}_{ij}|},
	\end{equation}
	where $\hat{r}_{ij}$ is the distance between lattice site $i_A$ and $j_B$, averaged over the spatial distribution of the respective particles around these sites\, \cite{carlesso2019} and $m_A$ and $m_B$ are the masses of the two particles. 
	\\ \noindent
	The assumption that gravity is quantum mechanical is implicit in this model of interaction. Because each component of the product state evolves with respect to a different Hamiltonian, the effective gravitational field between the two walks at a given time cannot be given by one field configuration. Hence, such an evolution assumes that the field itself is in a superposition of different configurations. A simplistic example explaining why such a field should be considered a quantum field is provided in Fig.\,\ref{modelsofgr}. With such a model of interaction, quantum information (like the amplitudes of each component) can be transmitted to another system. Such a field must be a quantum field because classical channels cannot transmit quantum information.
	\\ \\ \noindent
	The measure of distance in Eq. \,\eqref{hamiltonian} will depend on the kind of lattice we work on. But if we take the parallel distance between the two walks to be much greater than the lattice spacing as well as the region in which each walk spreads, it is safe to take the distance measure to be the Euclidean distance between the two sites. Furthermore, we assume that the background space-time on which the walk takes place is discretized at Plank scales. So the lattice parameter of the walk ($a$), is a multiple of the lattice parameter of the discrete space, which we take to be the Plank length $l_p$.   i.e. we define $a$ to be
	\begin{equation}
		a = |r(i_A)-r(i_A+1)|=N_d l_p.
	\end{equation}
	Here $r(i_A)$ is the position vector of the $i-th$ site of the walk $A$ and $N_d$ is the multiplication factor. So, the Euclidean distance between the sites $r(i_A)$ and $r(j_B)$ is given by
	\begin{equation}
		|\langle\hat{r}_{ij}\rangle|=N_d l_p\sqrt{L^2+|i_A-j_B|^2}\equiv N_dl_pd_{ij}.
	\end{equation}
	Here we have defined $d_{ij} = \sqrt{L^2+|i_A-j_B|^2}$. Similarly, the time taken between the two successive steps of the walk can be taken as a multiple of the Plank time $t_p$, $\delta t=N_t t_p$ with $N_t$ being the multiplicative factor. With this in mind, the component $\ket{i_A}\ket{j_B}$ in a walk of $t$ steps evolves under the Hamiltonian, given by Eq. \eqref{hamiltonian}, for the following amount of time
	\begin{equation}
		\Delta t_{ij}=N_tt_p\text{min}(t-|i|,t-|j|)-|\langle\hat{r}_{ij}\rangle|/c.
	\end{equation}
	The $-|\langle\hat{r}_{ij}\rangle|/c$ term ensures the locality of the gravitational interaction as it is the time taken for the carrier of interaction to travel the distance between the two lattice sites at the speed of light.
	\\
	As a result, the two-state system at time $t$ takes the following form due to the gravitational interaction,
	\begin{equation}\label{stateG}
		\begin{aligned}
			\ket{\Psi_G (t)} =  \sum_{i,j=-t}^{t}e^{-ig_{ij}(t)} & \Big ( \ket{i_A}\otimes(p^u_{i}\ket{\uparrow}+p^d_{i}\ket{\downarrow}) \Big ) \\
			\otimes \Big (& \ket{j_B}\otimes(q^u_{j}\ket{\uparrow}+q^d_{j}\ket{\downarrow})\Big ) .
		\end{aligned}
	\end{equation}
	Here $g_{ij}(t)$ is the phase acquired due to the gravitational potential between the site $i$ of state A and $j$ of state B.
	%\textcolor{lightgray}{ Since we are assuming that the states are performing the quantum walk on the lattice of space-time, we take the distance and time measures as some number times the Planck length $l_p$ and Planck time $t_p$ respectively. So the phase factor $g_{ij}(t)$ is given as:  }
	It is given by,
	\begin{equation}\label{phase}
		\begin{aligned}
			g_{ij}(t) = & -\dfrac{Gm_Am_B\Delta t_{ij}}{\hbar |\langle\hat{r}_{ij}\rangle|} \\
			=&-\dfrac{Gm_Am_B}{\hbar}\Big[\dfrac{N_tt_p\text{min}(t-|i|,t-|j|)}{N_d l_pd_{ij}}-\dfrac{1}{c}\Big] \\
			= & -\dfrac{m_Am_B}{{m_p}^2 }\Big[\dfrac{N_t\text{min}(t-|i|,t-|j|)}{N_d d_{ij}}-1\Big].
		\end{aligned}
	\end{equation}
	We have used the fact: $l_p/t_p=c$ and $\sqrt{c\hbar/G}=m_p$ being the Plank mass. We see that there is a constant phase term that can be dropped since entanglement depends only on the relative phases of the components.
	A discrete time one dimensional quantum walk with coin operation $C(\theta)$ of the form given in  Eq.\eqref{ucoin} is known to be a discretized model of the two component one dimensional Dirac equation\,\cite{srikant2010,chand2013,mallick2016dirac}. In the continuum limit of a DTQW, when the space parameter $a$ and the time parameter $\delta t$ tend to zero such that $a/\delta t=c$, c being the speed of light, the Hamiltonian of the walk resembles the Dirac Hamiltonian, given, we identify the mass of the Dirac particle with the coin parameters as $m=\hbar\sin(\theta)/\delta t$. Since we have assumed that the space is discretized into Plank length, and time is discretized into Plank time, we take $a=l_p$ and $\delta t=t_p$. So we approximate, 
	\begin{equation}
		\begin{aligned}
			\partial_t\psi \sim (\psi(t+t_p)-\psi(t))/t_p 
			\\
			\partial_z\psi \sim (\psi(z+l_p)-\psi(z))/l_p.
		\end{aligned}
	\end{equation}
	Using these substitutions for a walk in $z$ direction, we get the following dynamical equation,
	\begin{equation}\label{qwdirac}
		i\hbar\partial_t\psi=\big(-i\hbar c\sigma_z\partial_z+\dfrac{\hbar c}{l_p}\sigma_y\sin(\theta)\big)\psi,
	\end{equation}
	where $\sigma_z$ and $\sigma_y$ are the Pauli matrices. The two component Dirac equation is given by:
	\begin{equation}\label{diraceq}
		i\hbar\partial_t\psi=\big(-i\hbar c\hat{\alpha}\partial_z+\hat{\beta}mc^2\big)\psi.
	\end{equation}
	such that ${\hat{\alpha}}^2={\hat{\beta}}^2=1$ and $\{ \hat{\alpha},\hat{\beta} \}=0$.
	\color{black}
	Comparing the two Eqs.\,\eqref{qwdirac} and \eqref{diraceq}, we see that the mass of the Dirac state $\psi$ is proportional to $\sin(\theta)$,
	\begin{equation}\label{mass}
		m=m_p\sin{\theta}.
	\end{equation} 
	Plank mass $m_p$ is the proportionality constant. We use this definition of mass while evaluating the phase in Eq.\,\eqref{phase} to put in the mass of the quantum walkers. For a component $\ket{i_A}\ket{j_B}$, suppose $|j|>|i|$, then the associated phase will be given by
	\begin{equation}\label{grpotential}
		\begin{aligned}
			g_{ij}(t)= &-\dfrac{N_t}{N_d}\dfrac{m_Am_B(t-|j|)}{{m_p}^2d_{ij}}\\
			=&-\dfrac{N_t}{N_d}\dfrac{\sin(\theta_A)\sin(\theta_B)(t-|j|)}{d_{ij}}.
		\end{aligned}
	\end{equation}
	\\
	\noindent 
	With this expression we have shown that the phase acquired by each component due to gravity can be expressed solely in terms of the walk parameters.\\ 
	\color{black}
	\\
	\color{black}
	In the remaining part of this work, we will focus on the simulation of quantum walks under the given model of gravitational interaction. To make the computations unambiguous and less dependent on machine errors, we make the following two adjustments to the phase factor in Eq. \eqref{grpotential}: (1) We take $N_t=N_d$ which essentially means that at each step the hopping speed of the particle is $c$. This does not mean that the particle as a whole is moving at the speed of light because its velocity is determined by the group velocity of the wave function which is bounded above by $c$ times the cosine of $\theta$\,\cite{accwalk}. This is also a necessary condition for interpreting the quantum walk as a discrete version of Dirac Hamiltonian. (2) We use large values of $\theta$ ($\sim \pi/4$) which means that we are simulating walks with very large masses ($\sim m_p$). We do this because we run these simulations for a very small number of steps. One can in principle see measurable entanglement for masses $\sim 10^{-14}kg$ (for which superposition have been observed \cite{pino2018}) if the walks are performed for a very large number of steps.
	\\ \\
	\noindent
	\section{Entanglement}\label{four}
\begin{figure}[h]
		\centering
		\includegraphics[scale=0.35]{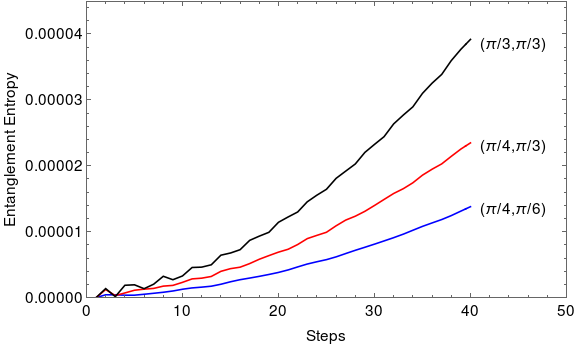}
		\caption{\small Entanglement entropy between two parallel 1D quantum walks under gravity for different mass pairs. Numbers in the bracket represent the mass parameters of the two states, $(\theta_A, \theta_B)$ where $\theta=\sin^{-1}(m/m_p)$). The initial spin states of the two walkers are (a) $\ket{\uparrow}$ and (b)$\ket{\downarrow}$, however we note that the initial spin states have no effect on EE.}
		\label{mi1}
	\end{figure}

	{\it Entanglement Entropy:}
	Entanglement entropy between two subsystems A and B is a von Neumann entropy of the density matrix reduced with respect to one of the systems. So if we trace out the spin and position space of the state A from the density matrix $\rho=\ket{\Psi_G}\bra{\Psi_G}$, we get the reduced density matrix $\rho_B$, and the entanglement entropy is then calculated as:
	\begin{equation}
		EE=-\sum_i\lambda_i\ln(\lambda_i) \hspace{2mm},
	\end{equation} 
	where $\lambda_i$ are the eigenvalues of $\rho_B$. This reduced density matrix takes the following form:
	\begin{equation}
		\rho_B=\sum_{j,k}\sum_{l}P_le^{-i(g_{il}-g_{jl})}\ket{j_B}\bra{k_B}\otimes \ket{s_j^B}\bra{s_k^B},
	\end{equation}
	where $P_l$ is the probability of the state A at site '$l$'. We can see this matrix as a perturbation to the pure density matrix $\rho_B^o=\sum_{jk}\ket{j}\bra{k}\otimes\ket{s_j^B}\bra{s_k^B}$, so that $\rho_B$ is a Hadamard product of the matrix $K_{ij}=\sum_{l}P_le^{-i(g_{il}-g_{jl})}$ with $\rho_B^o$, 
	\begin{equation}
		\rho_B=K\circ\rho_B^o.
	\end{equation} 
	The change in the eigen-spectrum due to a perturbation $\rho \rightarrow \rho+\delta\rho$ is given by,
	\begin{equation}
		{\lambda '}_i=\lambda_i+X_i^t\delta\rho X_i,
	\end{equation}
	where $X_i$ are the corresponding eigenvectors. Given that $\rho_B^o$ has one non-zero eigenvalue $\lambda_1=1$ with eigenvector $X_1=\sum_{j}\ket{j}\ket{s^B_j}$, the leading contribution to the new eigenvalue is given by:
	\begin{equation}\label{eigcorr}
		\delta\lambda_1'=\sum_{j,k,l}Q_jQ_kP_l(e^{-i(g_{il}-g_{jl})}-1).
	\end{equation}
	$\{Q_i\}$ are the probability distribution of the state B and $\{P_i\}$ are probability distribution of state A. A few more steps of calculation, under the approximation that  $L\gg t$, will show that the first order correction to the eigenvalue vanishes and the second order correction, which is a function of the second moment of the distributions of the two states provides the leading order contribution,
	\begin{equation}\label{eigcorrection}
		\begin{aligned}
			&	\delta^{(2)}\lambda_1'\propto \sin(\theta_A)^2\sin(\theta_B)^2\Big[ \sum_{l>i,j}Q_iQ_jP_l(t-l)^2 \\ &+ \sum_{l<i,j}Q_iQ_jP_l(t-i)(t-j) +\sum_{i>l>j}Q_iQ_jP_l(t-i)(t-l)\\&+\sum_{l>i,j}Q_iQ_jP_l(t-l)(t-j) \Big].
		\end{aligned}
	\end{equation}
	We measure the entanglement entropy for the state $\ket{\Psi_G}$ given by Eq.\,\eqref{stateG}.
	Based on the simulations run for different mass pairs of quantum walkers, we obtain EE between the two walkers as shown in Fig. \ref{mi1}. $\theta=\sin^{-1}(m/m_p)$ is the coin parameter (or the mass parameter). As the states evolve in the walk, EE between the two states typically increases quadratically with each time step. In addition, EE after a time $t$ is typically higher for higher $\theta$ values, till about $\theta \sim \pi/3$. The fact that the second moment of the quantum walks is a decreasing function of $\theta$ shows the significance of parameterizing mass of the walker in terms of the walk parameter (see Eq. \ref{mass}) (see Appendix for more discussions). Furthermore, EE is independent of the initial spin state of the walks, which also is an expected property as the even moments of a DTQW do not depend on the initial spin state \cite{konno2003}.
	\\
	%\textbf{Negativity:}\\
	{\it Negativity :}
	We calculate entanglement between the spatial degrees of freedom of the two interacting walks. By summing over the spin degrees of freedom of the product state $\ket{\Psi_G}$ we get the reduced density matrix $\rho^t$. The density matrix $\rho^t$ and can be calculated by first tracing out the spins of the subsystem $B$,
	
	\begin{equation}
		\rho^t_B=\bra{\uparrow_B}(\ket{\Psi_G}\bra{\Psi_G})\ket{\uparrow_B}+\bra{\downarrow_B}(\ket{\Psi_G}\bra{\Psi_G})\ket{\downarrow_B}
	\end{equation}
	followed by tracing out the spins of the subsystem $A$:
	\begin{equation}
		\rho^t=\bra{\uparrow_A}\rho_B^t\ket{\uparrow_A}+\bra{\downarrow_A}\rho_B^t\ket{\downarrow_A}.
	\end{equation}
	The resulting reduced density matrix $\rho^t$ is no longer in a pure state and hence entanglement entropy is not the right measure of entanglement between its two sub-systems (the spatial degrees of freedom of the walks). For mixed states, the Positive Partial Transpose (PPT) criterion gives a necessary condition to establish the separability of the density matrix. It says that if the partial transpose of the density matrix, with respect to any one of the sub-systems, is a positive density matrix, the total density matrix must have a separable form. A violation of this condition shows the presence of entanglement between the two subsystems. Negativity is a measure of entanglement derived from the PPT criterion.
	For a density matrix $\rho$, the negativity is the sum of the eigenvalues of its partial transpose, 
	%\begin{equation}
	$\mathcal{N}(\rho)=\sum_i\dfrac{|\lambda_i|-\lambda_i}{2}$,
	%\end{equation}\\
	where $\lambda_i$ are the eigenvalues of the partial transposed matrix $\rho^{\Gamma}$. Unlike some other measures of entanglement, negativity does not converge to entanglement entropy for pure states \cite{vidal2002} and is an entanglement monotone for $2\times 2$ and $2\times 3$  systems. However a positive value for negativity is a sufficient condition to establish entanglement in system\,\cite{peres1996,hor2001}.\\
	\begin{figure}[h]
		\centering	
		\includegraphics[scale=0.38]{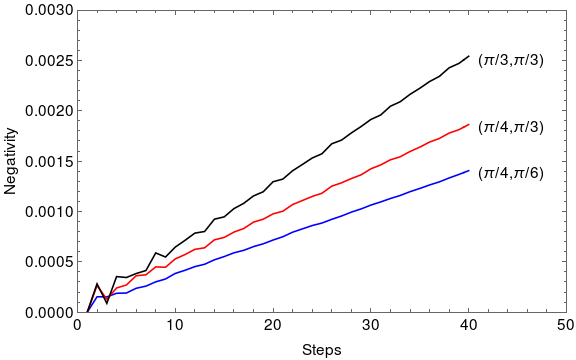}
		\caption{\small Negativity between two parallel 1-D quantum walks under Newtonian gravity, for different mass pairs. Numbers in the brackets are the mass parameters ($\theta=\sin^{-1}(m/m_p)$) of the two states. The initial spin states of the two walks are: (a)$\ket{\uparrow}$ and (b)$\ket{\downarrow}$.}
		\label{neg1}
	\end{figure}
	
\begin{figure}[h]
		\centering	
		\includegraphics[scale=0.22]{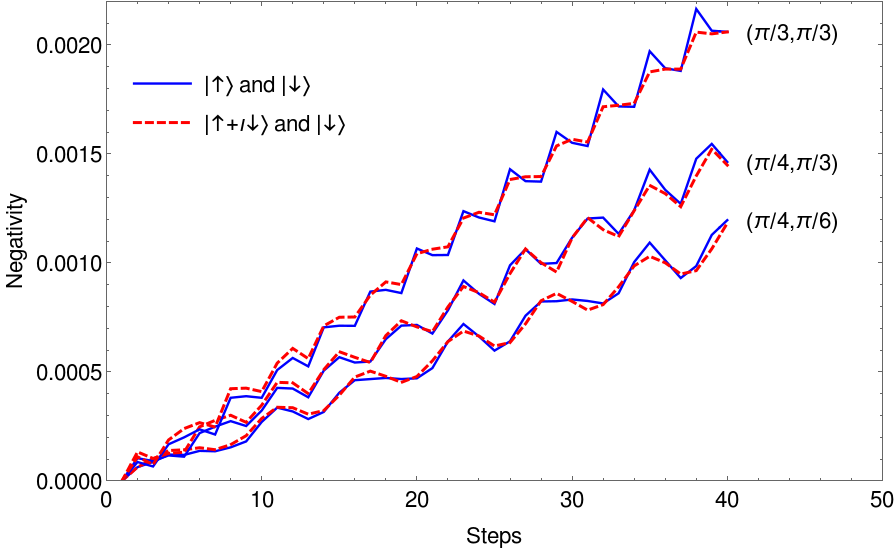}
		\caption{\small Negativity between two parallel 1-D quantum walks under Newtonian gravity with spin degrees of freedom traced out, for different mass pairs. Numbers in the brackets are the mass parameters ($\theta=\sin^{-1}(m/m_p)$) of the two states. Blue solid lines represent the walks with initial spin states (a) $\ket{\uparrow}$ and (b) $\ket{\downarrow}$. Red dashed lines represent the initial spin sates (a) $(\ket{\uparrow} -i\ket{\downarrow})/\sqrt{2}$  and (b) $\ket{\downarrow}$.}
		\label{tracedneg1}
	\end{figure}

	Negativity for three different mass pairs, parameterized by ($\theta_A,\theta_B$), is plotted in Fig.\,\ref{neg1} and Fig.\,\ref{tracedneg1} for the full system and after tracing out the spin degree of freedom. We see that negativity is independent of the initial spin state of the walks and increases linearly with the number of steps. Furthermore, just like the case of entanglement entropy, negativity for higher mass pairs is higher compared to the lower mass pairs after the same time steps. We see that the value of negativity after tracing out the spin degree of freedom follows the same trend as the original state $\ket{\Psi_G}$. However, entanglements for different pairs of initial states do not exactly overlap. This hints at the possibility that the negativity of a spin traced system also depends on the odd moments of the probability distributions.
	
\begin{figure}[h]
		\centering	
		\includegraphics[scale=0.22]{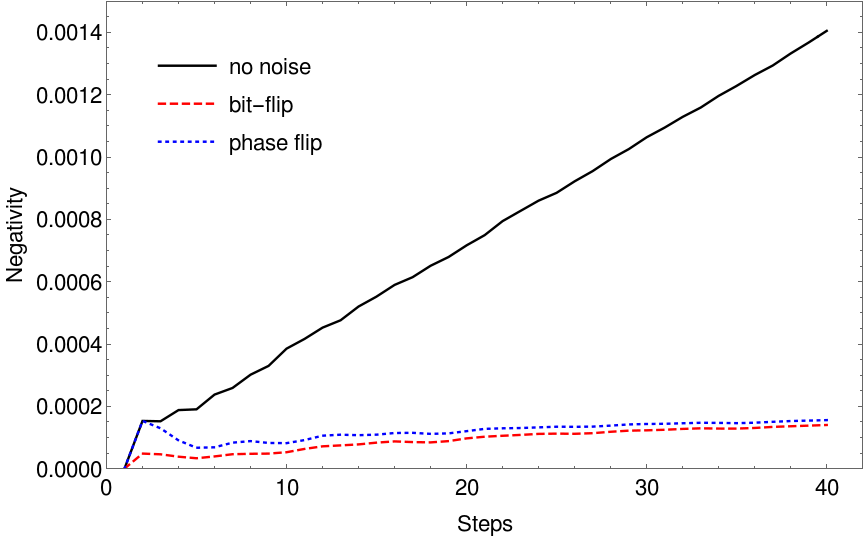}
		\caption{\small \textbf{Effect of Noise:} Black solid line shows negativity between two parallel 1-D quantum walks under Newtonian gravity with mass parameters ($\theta=\sin^{-1}(m/m_p)$) $\pi/4$ and $\pi/6$ with initial spin states $\ket{\downarrow}$ and $\ket{\uparrow}$ respectively. Red dashed line and blue dotted line shows negativity for the same setup, with a bit-flip and phase flip noise respectively introduced on the first state (with initial spin $\ket{\uparrow}$) with probability $p$= 0.02.}
		\label{noise1}
	\end{figure}

	\noindent
	{\it Effect of Noise:} Introduction of noise, like a simple bit flip ($\sigma_x$ gate) or a phase flip ($\sigma_z$ gate), introduced with a probability $p$ into a system reduces the effects of quantum interference in the dynamics of the system \cite{zureknoise,chandnoise}.  The ``two quantum walks" system separated by a large distance, that has been introduced in this work, has no interference among the walks themselves. But we have argued that there is a quantum mechanical channel between the two walks that is responsible for the gravitational interaction and as well as the entanglement generation between the two particles. Fig. \ref{noise1} shows the effect of the noise, applied on one of the walkers state, on the negativity between the two walkers. The reduction in entanglement is the indication of the fact that the local interactions between the ``carrier" of the gravitational interaction and the walk is a quantum mechanical interaction, proving that the carrier is a quantum mechanical state.
	\\ \\
	\section{Concluding remarks}\label{five}
	We have investigated the effects of a particular model of gravitational interactions on the discrete time quantum walks. We saw that the two walks get entangled with time if the interaction between them is mediated by a gravitational field that can treat each component of the superposition separately. This interaction obeys locality and since LOCC interactions cannot generate or increase entanglement, our results suggest a quantum carrier of gravitational interaction. This is also argued by the authors of\,\cite{ryanlocality}, wherein they suggest that it is the off-shell graviton that is exchanged between the two masses in superposition. The increase in entanglement with time explicitly presented in this work shows the importance of state's expansion in position space with superposition. The positive correlation between the rate of entanglement generation and coin parameter in the regimes where this parameter can be seen as a mass parameter of a Dirac particle, is a very interesting result that can further motivate the use of quantum systems to probe quantum nature of gravity. Finally we show that the introduction of noise in one of the walks results in a reduction of entanglement which further strengthens the argument that the interaction between the walk and the gravitational channel is indeed quantum mechanical.
	\\ 
	Although we have shown the entanglement generation for high values of mass parameters ($\sim m_p$) for illustration, the results hold for small masses for which superposition has been seen in a lab. For smaller masses ($\sim 10^{-14}kg$) but for a much larger number of steps, the entanglement might be detectable in a lab. In addition to this, we have the freedom to extend the study to higher dimensional walks where entanglement generation rate will be comparatively faster.
	\\
	We acknowledge the fact that one does not know if gravitational interaction does indeed remain Newtonian at the scales we are studying them. Newton's law of gravity has been tested reliably only at the scales of the solar system. At higher scales, general relativity takes over and there is no reason to believe that it should hold at microscopic scales too \cite{murata2015}. However, a deeper question still remains unanswered, which is: what is the \textit{nature} of gravitational interaction for states in a superposition of positions  \cite{carlesso2019, ryanlocality}? Gravity could still be quantum mechanical and Newtonian but act in a way that is different from our assumptions in this paper. With access to controllable quantum systems and a carefully designed experiments could lead to some answer in coming days.
	\\ \\
	\clearpage
	\appendix
	\section{Variation of entanglement with respect to the coin parameters}
\begin{figure}[h!]
		\begin{subfigure}[b]{0.2\textwidth}
			\includegraphics[scale=0.4]{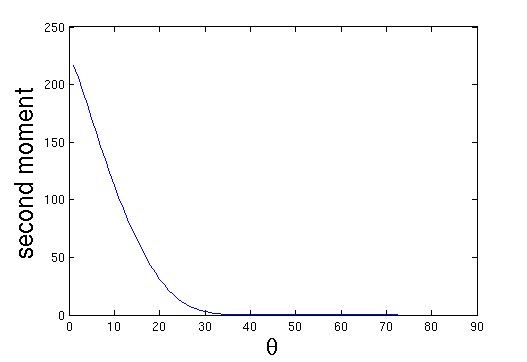}
			\caption{}
			\label{amoment}
		\end{subfigure}\\
		\begin{subfigure}[b]{0.2\textwidth}
			\includegraphics[scale=0.4]{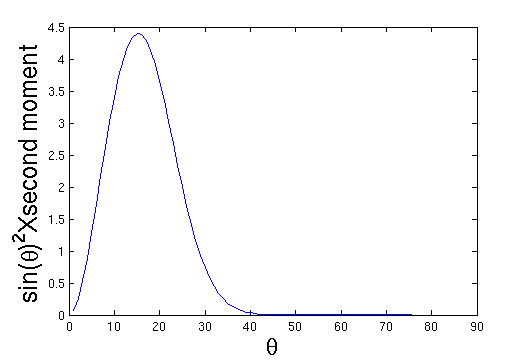}
			\caption{}
			\label{bmoment}
		\end{subfigure}
		\caption{(a) shows the variation of the second moment about the mean of the probability distribution of a quantum walk, after 15 steps of the walk with respect to the coin parameter. (b) shows the effect of the $\sin(\theta)^2$ factor multiplying the second moment. This shows that the positive correlation between the mass parameter and entanglement generation can be seen only for smaller values of the parameter.}
	\end{figure}

\begin{figure}
		\begin{subfigure}[b]{0.2\textwidth}
			\includegraphics[scale=0.15]{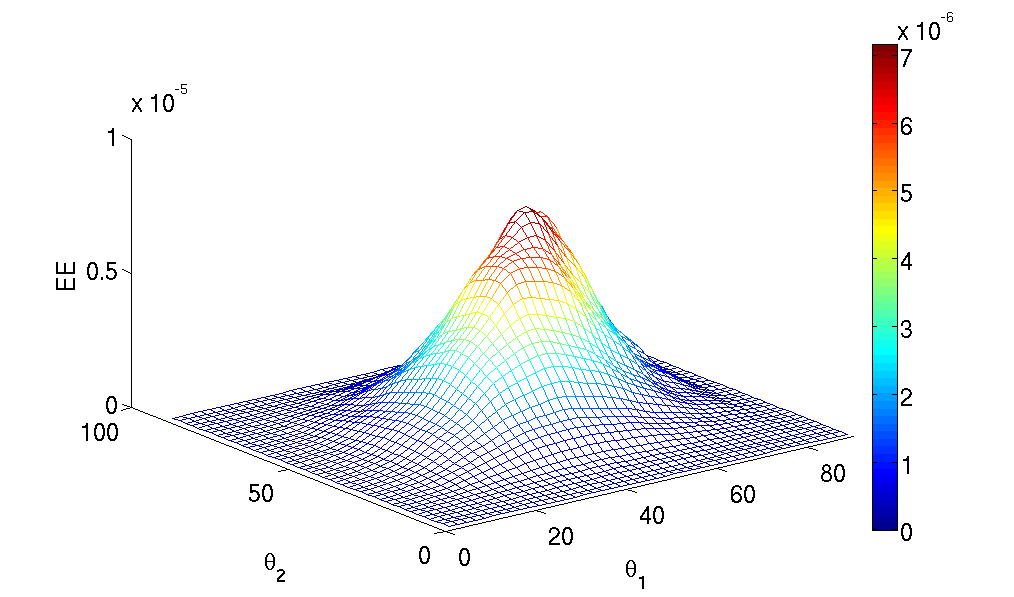}
			\caption{}
		\end{subfigure}\\
		\begin{subfigure}[b]{0.2\textwidth}
			\includegraphics[scale=0.15]{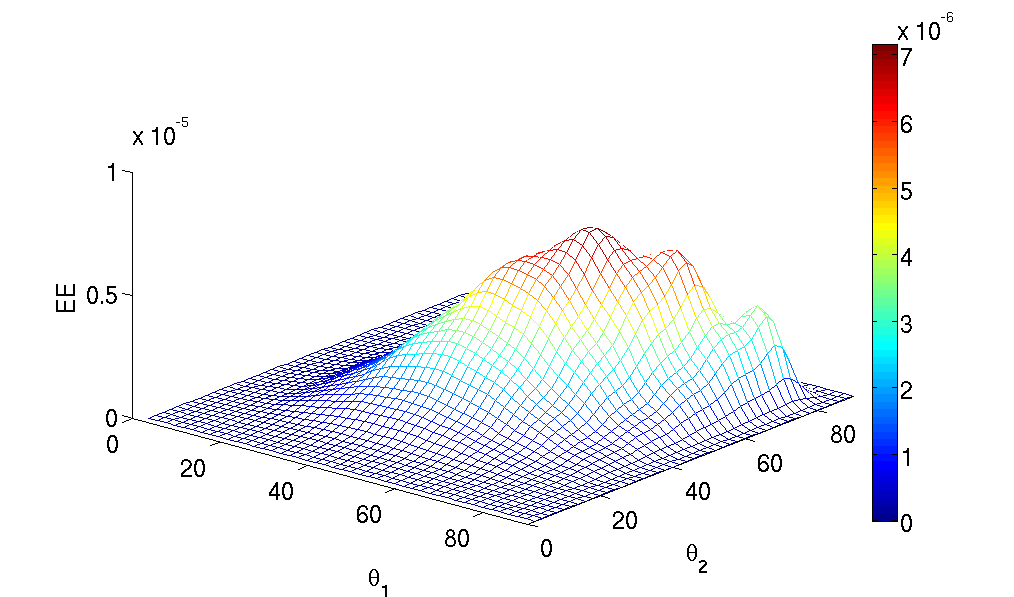}
			\caption{}
		\end{subfigure}\\
		\begin{subfigure}[b]{0.2\textwidth}
			\includegraphics[scale=0.15]{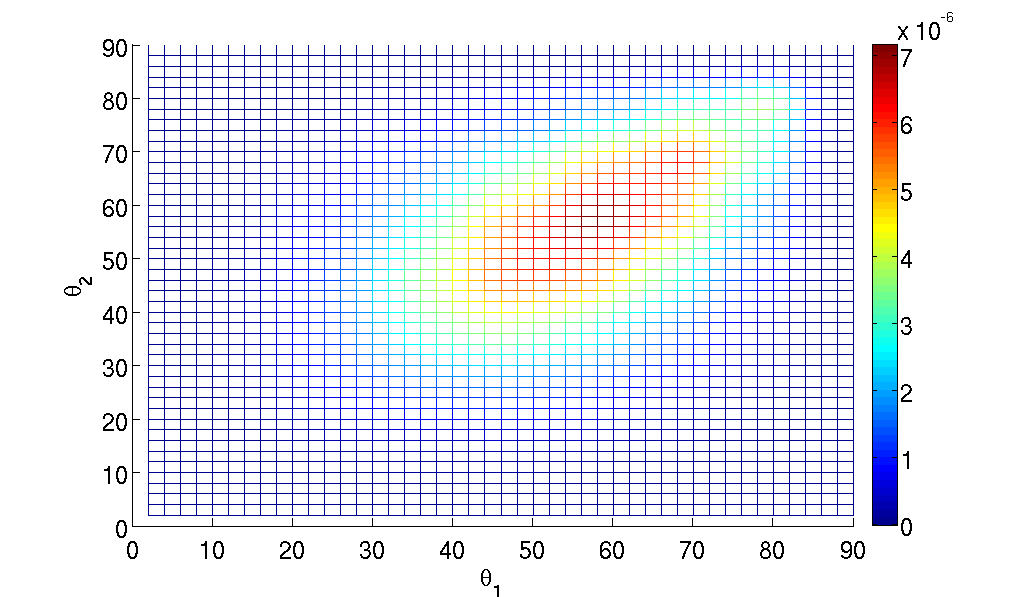}
			\caption{}
		\end{subfigure}
		
		\caption{Variation of entanglement entropy with the $\theta$ parameters after 15 steps of the walk. Initial spin states are $\ket{\uparrow}$ and $\ket{\downarrow}$. All the figures are the same graph from different orientations.}
		\label{eetheta}
	\end{figure}

\begin{figure}
		\begin{subfigure}[b]{0.2\textwidth}
			\includegraphics[scale=0.15]{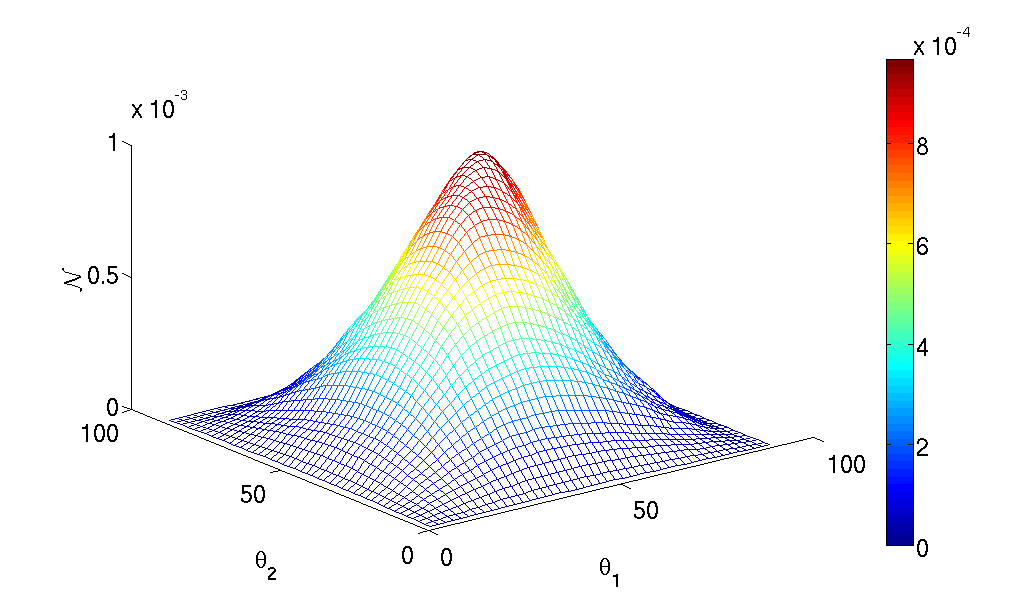}
			\caption{}
		\end{subfigure}\\
		\begin{subfigure}[b]{0.2\textwidth}
			\includegraphics[scale=0.15]{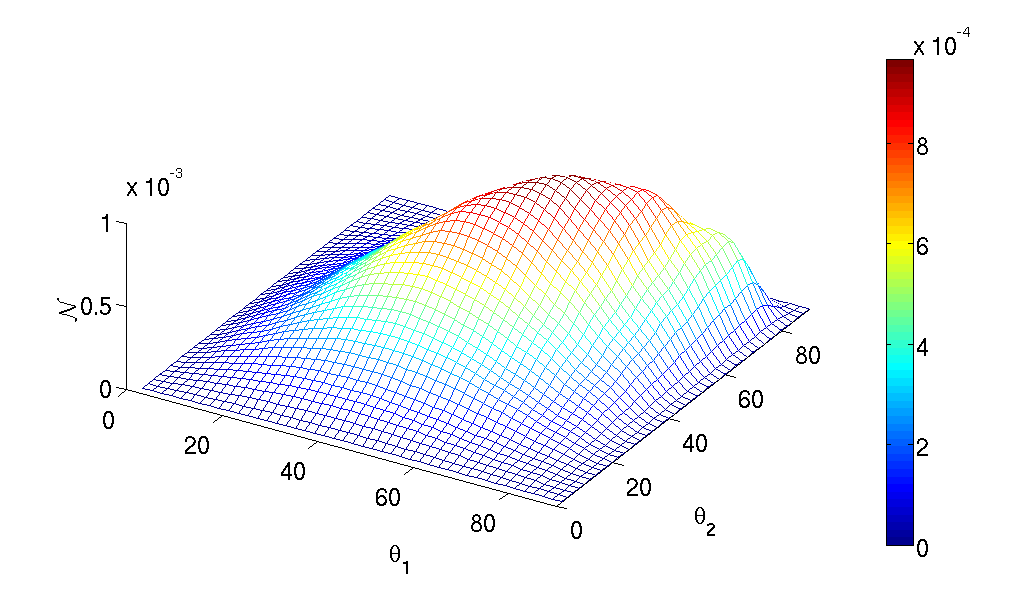}
			\caption{}
		\end{subfigure}\\
		\begin{subfigure}[b]{0.2\textwidth}
			\includegraphics[scale=0.15]{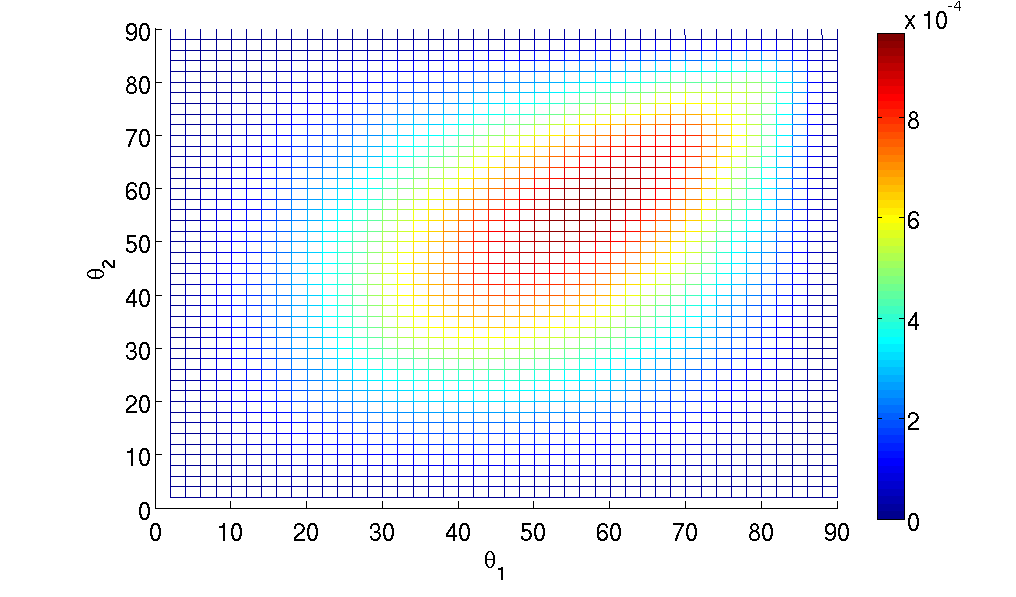}
			\caption{}
		\end{subfigure}
		\caption{Variation of negativity with the $\theta$ parameters after 15 steps of the walk. Initial spin states are $\ket{\uparrow}$ and $\ket{\downarrow}$. All the figures are the same graph from different orientations.} 
		\label{negtheta}
	\end{figure}

	\noindent
	As discussed in the main text, the analytical solution for the correction to the eigenvalue (Eq. \eqref{eigcorr}) suggest that the entanglement should initially increase with $\theta$ due to the $\sin(\theta)^2$ term until the point when the term that is a function of second moments of the distributions takes over. Fig. \ref{amoment} shows how the second moment varies with the coin parameter for a given walk and Fig. \ref{bmoment} shows the variation of the product of $\sin(\theta)$ with the second moment. In Fig. \ref{eetheta} and Fig. \ref{negtheta} we show the variation of entanglement entropy and negativity respectively with respect to the coin parameters $\theta_1$ and $\theta_2$. We see that the entanglement increases until about $\theta \sim \pi/3$ and decreases after that. The substitution of Eq. \eqref{mass} thus gives us a positive correlation between mass and entanglement for small masses.
	
	\clearpage
	%===================================================================
	
	\end{document}